\documentclass[10pt,a4paper]{article} 
\usepackage{graphicx}
\usepackage{amsfonts,amsmath,amssymb}
\usepackage{hyperref}
\usepackage{cite}
\newcommand{\bse}{\begin{subequations}}
\newcommand{\ese}{\end{subequations}}

\begin{document}
IPM/P-2017/007
\begin{center}
{ \Large{\textbf{Far-from-equilibrium initial conditions probed by a \\\vspace{.2cm}nonlocal observable}}} 
\end{center}
\vspace*{0.3cm}
\begin{center}
{\bf L. Shahkarami$^{a,d,1}$, H. Ebrahim$^{b,d,2}$, M. Ali-Akbari$^{c,3}$, F. Charmchi$^{e,4}$}\\
\vspace*{0.3cm}
{\it {${}^a$School of Physics, Damghan University, Damghan, 41167-36716, Iran}}\\
{\it {${}^b$Department of Physics, University of Tehran, North Karegar Ave., Tehran 14395-547, Iran}} \\
{\it {${}^c$Department of Physics, Shahid Beheshti University G.C., Evin, Tehran 19839, Iran}}\\
{\it {${}^d$School of Physics, Institute for Research in Fundamental Sciences (IPM),
P.O.Box 19395-5531, Tehran, Iran}} \\
{\it {${}^e$School of Particles and Accelerators, Institute for Research in Fundamental Sciences (IPM),
P.O.Box 19395-5531, Tehran, Iran}}  \\
{\it  {${}^1$l.shahkarami@du.ac.ir}, {${}^2$hebrahim@ut.ac.ir}, {${}^3$$m_{-}$aliakbari@sbu.ac.ir}, {${}^4$charmchi@ipm.ir}  }
\end{center}

\begin{abstract}
Using the gauge/gravity duality, we investigate the evolution of an out-of-equilibrium strongly-coupled plasma from the viewpoint of the two-point function of scalar gauge-invariant operators with large conformal dimension.
This system is out of equilibrium due to the presence of anisotropy and/or a massive scalar field.
Considering various functions for the initial anisotropy and scalar field, we conclude that the effect of the anisotropy on the evolution of the two-point function is considerably more than the effect of the scalar field.
We also show that the ordering of the equilibration time of the one-point function for the non-probe scalar field and the correlation function between two points with a fixed separation can be reversed by changing the initial configuration of the plasma, when the system is out of the equilibrium due to the presence of at least two different sources like our problem.
In addition, we find the equilibration time of the two-point function to be linearly increasing with respect to the separation of the two points with a fixed slope, regardless of the initial configuration that we start with.
Finally we observe that, for larger separations the geodesic connecting two points on the boundary crosses the event horizon after it has reached its final equilibrium value, meaning that the two-point function can probe behind the event horizon.
\end{abstract}


\section{Introduction}

Understanding the out-of-equilibrium dynamics of a strongly-coupled gauge theory has attracted much interest in the past decade.    
One motivation for this comes from the results obtained in the ultra-relativistic heavy ion collisions at the Brookhaven RHIC and CERN LHC.  
The experimental observations, such as the elliptic flow, indicate that the hot and dense medium, created at the early moments of the collisions, is more similar to a perfect fluid with a small entropy-normalized viscosity than a collection of the individual partons \cite{Shuryak4,Shuryak5}.  
This medium called Quark-Gluon Plasma (QGP) is a strongly-coupled phase of QCD which is highly out of equilibrium and takes a very short time of order 1 $fm/c$ to reach a thermalized state \cite{Heinz5}.  
A challenging and still unresolved question is how a strongly-coupled out-of-equilibrium gauge theory evolves towards its thermal equilibrium.

A powerful tool to study the dynamics of such systems is the gauge/gravity duality \cite{Maldacena1,Maldacena2,Maldacena3,Solana} which solves the previously intractable problems using the classical gravitational dynamics of the asymptotically anti-de Sitter spacetimes.
Plenty of work has been done on the early-time dynamics and subsequent thermalization in the past decade using various holographic techniques \cite{thermal1,thermal2,chesler1,chesler2,thermal3,thermal7,Heller1,thermal4,Heller2,thermal5,thermal6,cheslernum,Ishii,us1,us,Kaviani,bitaghsir}. 
In particular, one way for investigating the dynamics of strongly-coupled far-from-equilibrium systems proposed in \cite{chesler1,chesler2,cheslernum} is to drive the field theory out of equilibrium by distorting the metric for a finite amount of time in such a way that an anisotropy is created between the longitudinal and transverse directions.
By solving the Einstein equations using the elegant method of characteristic formulation, they were able to study the isotropization of the system.
Another way for investigating the early-time dynamics of far-from-equilibrium systems is to start from an out-of-equilibrium plasma, in the absence of external sources, and then allow it to evolve and eventually reach its thermal equilibrium \cite{Heller1,Heller2}.
In this approach the boundary metric is flat and time-independent, and the energy density is conserved.
If the initial state is anisotropic, the pressure in the longitudinal and transverse directions will evolve in time and hence we are able to study the isotropization process.

One can also add an operator to the field theory and observe the time evolution of its one-point function for studying the thermalization process.
In \cite{us} we followed the second approach considering an initially anisotropic geometry interacting with a non-probe massive scalar field.
We supposed that the expectation value of the non-probe scalar field at equilibrium is zero.
Computing the pressure anisotropy and expectation value of the non-probe scalar field as time proceeds, we were able to study the process of approaching the equilibrium.
For our purposes in most of the calculations we employed a Gaussian profile with tractable parameters as the initial function for both the non-probe scalar field and anisotropy function.
By changing the parameters, we concluded that depending on the configuration of the initial plasma, various ordering of relaxation times, i.e.\;isotropization, thermalization and equilibration times, are possible.
As introduced in \cite{us}, by these three time-scales we mean the time at which longitudinal and transverse pressures become equal, the entropy production ends, and the expectation value of the non-probe scalar field reaches its equilibrium, respectively, within the numerical precision chosen.

Extended objects such as correlation function of two local gauge-invariant operators, expectation values of Wilson loops and entanglement entropy are more sensible to the details of the thermalization process than local probes \cite{balasubramanian1,balasubramanian2,galante,pedrazanon,nonlocal1,nonlocal2,nonlocal3,nonlocal4}.
Providing an elegant geometric way to compute these observables, AdS/CFT gives an explanation for this statement; nonlocal probes penetrate deeper into the bulk, hence probe a wider range of energies in the boundary field theory.
In fact, since nonlocal probes distinguish different scales, they can be used for considering the contribution of different scales of energy in the thermalization process.

Our purpose in the present paper is to investigate the thermalization process by studying the time evolution of both local and nonlocal observables.  
To that purpose, we consider the solution obtained in \cite{us} as an out-of-equilibrium background.
On the gravity side, this system is an initially anisotropic geometry interacting with a non-probe scalar field dual to a scalar operator of conformal dimension $\Delta=3$.
Using numerical calculation we find the time-evolution of the expectation value of this scalar operator.
Moreover, we calculate the equal-time two-point correlation function for another scalar operator which has a large conformal dimension $\Delta \gg 1$ and its backreaction to  the field theory is not taken into account.  
According to the AdS/CFT dictionary, to obtain the evolution of the two-point function in the field theory, we compute the time evolution of the geodesic length connecting the two points where the local operators are inserted at the boundary.
Changing the initial configuration of the system, we would be able to study the evolution of the two-point function for different time ordering of the relaxation time-scales.
The main difference between our problem with the problems in \cite{nonlocal2,nonlocal4} is the presence of the non-probe scalar field beside the anisotropy.
This enables us to study the effect of being out of equilibrium due to more than one source.

The main results we obtain in this paper can be summarized as follows.
\begin{itemize}
\item
We consider the evolution of several initial configurations, one of them is anisotropic and coupled to a non-probe scalar field and the other ones are isotropic with non-probe scalar field and anisotropic without any scalar field.
They are chosen to have the same functions with the same parameters so that they thermalize at considerably similar times, i.e.\;the area of the event horizon for them reach their final values at similar times.
Comparing the evolution of their two-point functions,
we found the effect of the scalar operator on the two-point function to be much less than the anisotropy.
The non-probe scalar field has considerable effects on the two-point function evolution, only when it is chosen to be very far from equilibrium, i.e.\;its initial function concentrate near the event horizon.
\item
We show that the ordering of the equilibration of the non-probe scalar one-point function and the probe two-point function can be reversed, by changing the parameters of the initial configuration and this is possible only when there are more than one source for deriving the system out of equilibrium.
We should mention here that in \cite{us} we concluded that the ordering of the isotropization, thermalization and equilibration times can be changed by changing the initial configuration.
Notice that in the case where there is only one cause for being out of equilibrium, no matter how we change the initial configuration of the system, the ordering of the time-scales remains unchanged.
In general, for an out-of-equilibrium system, time ordering of the equilibration of different probes depends on the initial configuration.
\item
We also find that the equilibration time of the two-point function increases linearly with the separation of the points, for large enough separations, and the slope of this growth is independent of the initial configuration of the system.
\item
Finally, we show that after the event horizon approaches its equilibrium, the geodesics for large separations on the boundary crosses the event horizon.
\end{itemize}

The paper is organised as follows.
In the next section, we briefly review the calculation of the evolving background metric.
Interested readers are referred to \cite{us} and  the above-mentioned original papers of the characteristic formulation, for more details.
We explain the strategy we use for the calculation of the geodesic length in section 3.
Finally, in section 4 we present the results of the numerical calculations for the geodesic length in some graphs and discuss about the thermalization time of the two-point function in different situations.
\section{Evolution of an anisotropic spacetime with an scalar field}
This section is devoted to a short review of the problem solved in \cite{us}. 
The metric ansatz for an anisotropic asymptotically $\mathrm{AdS}_5$ spacetime in the generalized Eddington-Finkelstein coordinates can be written in the following form 
\begin{align}\label{metric}
ds_5^2&=2 dr dv-A(v,r)dv^2+\Sigma(v,r)^2 \left(e^{-2B(v,r)} dx_L^2+e^{B(v,r)}d\bold{x}_T^2\right).
\end{align}
This metic is translationally invariant in the spatial coordinates of the boundary which is
 located at $r \to \infty$. 
$v$ is the time coordinate in the bulk and is the boundary time $t$ as $r \to \infty$.
The function $B(v,r)$ introduces an anisotropy between the  longitudinal (${x}_L$) and transverse ($\bold{x}_T$) directions of the boundary.
We study the Einstein's general relativity minimally coupled to a massive scalar field with $m^2=-3$ which is dual to a fermionic mass operator of conformal dimension $\Delta=3$.
By varying the action of the gravity side with respect to the metric and the scalar field and then inserting the ansatz (\ref{metric}) into the resulting equations, we obtain the following equations in terms of the unknown functions $A(v,r)$, $B(v,r)$, $\Sigma(v,r)$ and $\phi(v,r)$,  
\bse\label{EOM}\begin{align}
\label{1} 0&= \Sigma\partial_r(\dot{\Sigma})+2\dot{\Sigma}\partial_r\Sigma
-2\Sigma^2+\frac{1}{12}m^2\phi^2\Sigma^2,\\
\label{2} 0&=2\partial_r(\dot{\phi})+\frac{3\partial_r\Sigma}{\Sigma}\dot{\phi}
+\frac{3\partial_r\phi}{\Sigma}\dot{\Sigma}-m^2\phi,\\
\label{3}  0&=2\partial_r (\dot{B})+\frac{3\partial_r \Sigma}{ \Sigma}\dot{B}+\frac{3\dot{\Sigma}}{\Sigma}\partial_r B,\\
\label{4} 0&=\partial_r^2A-\frac{12}{\Sigma^2}\dot{\Sigma}\partial_r\Sigma+3\dot{B}\partial_r B
+4+\dot{\phi}\partial_r\phi-\frac{1}{6} m^2\phi^2,\\
\label{5} 0&=\ddot{\Sigma}-\frac{1}{2} \partial_r A \dot{\Sigma}+\frac{1}{6} \Sigma(3\dot{B}^2+\dot{\phi}^2),\\
\label{6} 0&=\partial_r^2\Sigma+\frac{1}{6}\Sigma\left(3(\partial_r B)^2+(\partial_r\phi)^2\right),
\end{align}\ese
where $ \dot{h} \equiv \partial_v h+\frac{1}{2}A\partial_r h$ denotes derivative of any function $h(v,r)$ along the radial outgoing null geodesics.

To solve this set of equations, we need to start from an appropriate initial configuration which involves two functions for $B(0,r)$ and $\phi(0,r)$ plus the value of the energy density.
The field theory would eventually approach an anisotropic plasma with no scalar field.
Also, we should impose convenient conditions at the boundary, i.e.\;$r\to \infty$.
The bulk metric should approach that of the field theory, which is the 4-dimensional Minkowski metric and the source of the non-probe scalar field is considered to be zero. 
Having the functions $B(v,r)$ and $\phi(v,r)$ at any time $v$, we can solve the 
Eqs.\,(\ref{EOM}) one by one, each one as an ordinary differential equation of $r$, with the aid of the pseudo-spectral method.
Then using an adequate method such as Runge-Kutta we can proceed to the next time step $v+\delta v$.
In most cases we choose the following Gaussian profile with tractable parameters ${\cal A}$ and $\beta$ controlling the amplitude and the position of the pulse, respectively
\begin{align}\label{prof}
f(0,r)={\cal A}\  {r^{-4}} \  e^{{-100}\big(\frac{1}{r}-\beta\big)^2}.
\end{align}

In order to study the thermalization process in such a background, one can monitor its dynamics by the use of a variety of observables of the field theory.
Local probes, such as the expectation value of the energy-momentum tensor and the non-probe scalar field response, provide valuable information about how the system makes progress towards equilibration. 
However, by being sensitive to the form of the geometry in the deep IR, nonlocal observables are able to give much more insight into the problem. 
In what follows, we study the evolution of the two-point correlation function of a local gauge-invariant operator as a nonlocal probe in this background.
Next section is devoted to the calculation of this correlation function.
\section{Two-point correlation function}
We want to calculate the equal-time two-point function for a local scalar operator ${\cal O}(t,\bold{x})$ of conformal dimension $\Delta$ in the aforementioned time-dependent background.
For quantum field theories with dimensions higher than two, there is no known technique to calculate the two-point function, analytically.
Thus, we have to resort to approximate and/or numerical schemes, specially when we are dealing with non-equilibrium backgrounds.
Under special circumstances, we can evaluate it using AdS/CFT correspondence.
For space-like trajectories and large conformal dimension $\Delta \gg 1$ the so-called geodesic approximation can be applied to obtain the following relation for the equal-time two-point function \cite{geodesic1,geodesic2}
\begin{align}
\left \langle {\cal O}(t,\bold{x}){\cal O}(t,\bold{x'})\right \rangle\approx \sum_{\mathrm{geodesics}} e^{-\Delta {\cal L}},
\end{align}
where ${\cal L}$ is the real length of the space-like geodesics connecting the boundary points $(t,\bold{x})$ and $(t,\bold{x'})$. 
To leading order, which is of our interest, the only contribution to the two-point correlation function comes from the smallest value of ${\cal L}$.

This geometrical description explains why the two-point function and other nonlocal probes provide more information than local probes about the evolution of out-of-equilibrium systems.
For small enough separations of the boundary points the associated geodesic does not penetrate very much into the bulk and therefore it will thermalize very fast.
As we enlarge the separation of the points, the geodesic will penetrate deeper into the bulk and the thermalization of the two points occurs later.
This leads us to the conclusion that the ultraviolet degrees of freedom thermalizes first.
\subsection{Numerical calculation of the geodesic length}
Now, we describe the strategy we use to calculate the length of the space-like geodesics between the two equal-time points $(t_0,-l/2)$ and $(t_0,l/2)$ \cite{balasubramanian2,galante,nonlocal4}. 
These two points can be separated either in the longitudinal direction $x_L$ or in one of the transverse directions ${\bf x}_T$ which are the same due to the rotational symmetry in the transverse plane.
We take the coordinate of the separation of the points as the parameter of the geodesic between the points and for brevity of notation we denote it simply as $x$.
Therefore, the geodesic will be obtained by determining the functions $v(x)$ and $r(x)$.
The following boundary conditions should be fulfilled by these functions
\begin{align}
r(-l/2)=r(l/2)=r_0,~v(-l/2)=v(l/2)=t_0,
\end{align}
where $r_0$ is a bulk cut-off close to the boundary. 
Also, since $v(x)$ and $r(x)$ are symmetric under reflection $x \to -x$, they satisfy the following conditions at the midpoint $x=0$
\begin{align}\label{conds0}
r'(0)=v'(0)=0,~r(0)=r_*,~v(0)=v_*,
\end{align}  

The geodesic length between two points on the boundary is obtained by extremizing the length of the curves connecting those two points, which is defined as follows for the bulk metric (\ref{metric})
\begin{align}\label{lagrang}
{\cal L}=\int dx \left[-A(v,r)v'(x)^2+2 v'(x) r'(x) +{\tilde \Sigma}(v,r)\right]^{1/2},
\end{align}  
where ${\tilde \Sigma}(v,r) \equiv \Sigma(v,r)^2 e^{B(v,r)}$.
We can interpret the integrand of this equation as a Lagrangian and since it is not dependent explicitly on $x$, there exists a conserved quantity which can be expressed as
\begin{align}\label{hamilton}
{\tilde \Sigma}(v_*,r_*)^{1/2}=\frac{{\tilde \Sigma}(v,r)}{\left[-A(v,r)v'(x)^2+2 v'(x) r'(x) +{\tilde \Sigma}(v,r)\right]^{1/2}}.
\end{align}  
The left-hand side has been written using the conditions (\ref{conds0}).
By inserting this relation into the Euler-Lagrange equations obtained by variation of Eq.\,(\ref{lagrang}) we obtain two second order differential equations for $v(x)$ and $r(x)$ as follows 
\bse\label{equs}\begin{align}
\label{11}&A(v,r)v''(x)-r''(x)+\left[\frac{\partial_v {\tilde \Sigma}(v,r)}{{\tilde \Sigma}(v,r)}-A(v,r) \frac{\partial_r {\tilde \Sigma}(v,r)}{{\tilde \Sigma}(v,r)} +\partial_r A(v,r)\right]r'(x)v'(x)\nonumber\\
&~~+\left[\frac{1}{2}\partial_v A(v,r)-A(v,r)\frac{\partial_v {\tilde \Sigma}(v,r)}{{\tilde \Sigma}(v,r)} \right]v'(x)^2+\frac{\partial_r {\tilde \Sigma}(v,r)}{{\tilde \Sigma}(v,r)} r'(x)^2+\frac{1}{2}\partial_v {\tilde \Sigma}(v,r)=0,\\
\label{22}&v''(x)-\frac{\partial_r {\tilde \Sigma}(v,r)}{{\tilde \Sigma}(v,r)}r'(x)v'(x)+\left[\frac{1}{2}\partial_r A(v,r)-\frac{\partial_v {\tilde \Sigma}(v,r)}{{\tilde \Sigma}(v,r)} \right]v'(x)^2-\frac{1}{2}\partial_r {\tilde \Sigma}(v,r)=0.
\end{align}\ese   
The functions $v(x)$ and $r(x)$ are determined by solving these equations numerically for   given values of $v_*$ and $r_*$.
Substituting the profile of the geodesic into Eq.\,(\ref{lagrang}), we can determine the corresponding geodesic length as
\begin{align}\label{length}
{\cal L}=\int_{-l/2}^{l/2} dx \frac{{\tilde \Sigma}(v,r)}{\sqrt{{\tilde \Sigma}(v_*,r_*)}}.
\end{align}  
We regularize this expression by subtraction of the divergent part of the geodesic length computed in pure AdS and find the regularized geodesic length ${\cal L}_{\mathrm{reg}}$.
\section{Results and concluding remarks}
By solving the differential Eqs.\,(\ref{equs}) with the relations (\ref{conds0}) as initial conditions, we are able to find the length of the geodesic corresponding to each pair of $v_*$ and $r_*$.
We have computed the geodesic length as a function of time for several initial configurations by changing the parameters $\beta_{B,\phi}$ and ${\cal A}_{B,\phi}$ appearing in the profile (\ref{prof}) for the anisotropy function and the non-probe scalar field.
Here we report on some conclusions which can be drawn by comparing different initial configurations.

We first set the notation and give some definitions used in this section.
In what follows $L(t)$ is the rescaled geodesic length defined as
\begin{align}\label{rel}
L(t)=\frac{{\cal L}_{\mathrm{reg}}(t)-{\cal L}_{\mathrm{reg}}^{\mathrm{th}}}{l},
\end{align}  
where ${\cal L}_{\mathrm{reg}}^{\mathrm{th}}$ is the equilibrium value of the regularized geodesic length.
We define the equilibration time of the 2-point function ($t_{\mathrm{L}}$) as the time after which $|L(t)|$ is lower than some small number $\epsilon$.
Moreover, as defined in \cite{us}, the equilibration time of the scalar one-point function ($t_{\mathrm{eq}}$) is the time after which $|{\cal E}^{3/2}\phi_2(t)|<\epsilon$ (where ${\cal E}$ is the energy density) and 
 the thermalization time ($t_{\mathrm{th}}$) is the time after which the entropy production ends or equivalently the event horizon reaches its equilibrium, and can be determined by $\left| \frac{r_{EH}(t_{\mathrm{th}})-\pi T}{\pi T}\right |<\epsilon$, where $r_{EH}(t)$ and $T$ are the event horizon radius and the final temperature of the plasma, respectively.
Notice that $\epsilon$ is a small number that sets the precision for determining the time-scales and is chosen to be $0.005$ in most of the cases here.
The results that are reported below have been checked for different precisions and we found them to be independent of the choice of this number.
Furthermore, in all figures the boundary time $t$ and separation length $l$ have been rescaled in units of the temperature of the final static black hole, i.e.\;$T=1/\pi$.

\begin{figure}[h]
\begin{center}
\includegraphics[width=6.cm]{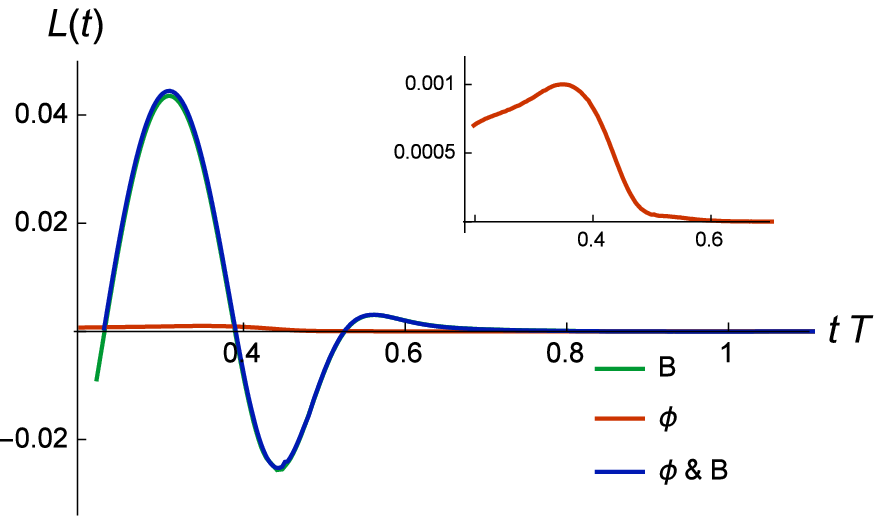}\hspace{.7cm}
\includegraphics[width=6.cm]{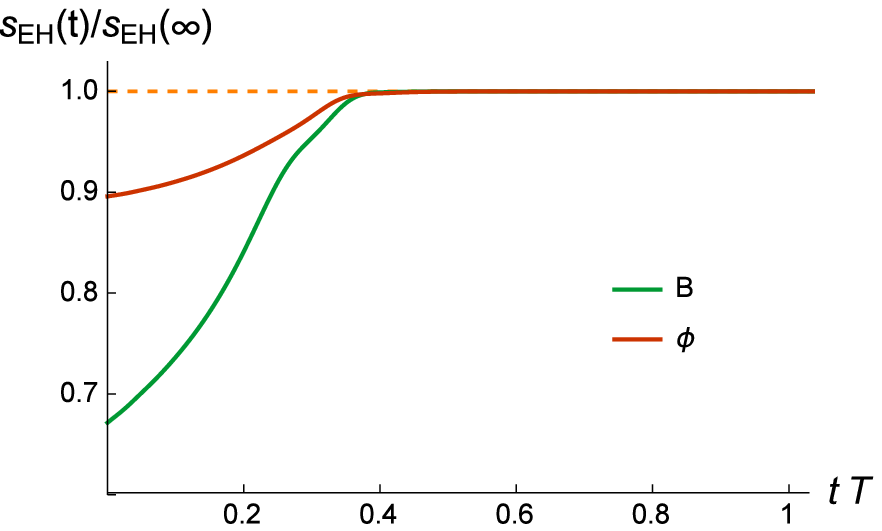}
\end{center}
\caption{\footnotesize 
Left graph: geodesic length versus time for the transverse separation $lT=0.31831$ and right graph: the time evolution of the entropy density.
Green, red and blue lines are the graphs for the initial configurations with $\phi = 0$, $B=0$ and the case with both functions present, respectively.
The functions are in the form of Eq.\,(\ref{prof}) with parameters  
$\beta_B=0.5$, ${\cal A}_B=1.5$, $\beta_{\phi}=0.5$ and ${\cal A}_{\phi}=1.5$.
}
\label{L}
\end{figure} 

\begin{figure}[htp]
\begin{center}
\includegraphics[width=7.cm]{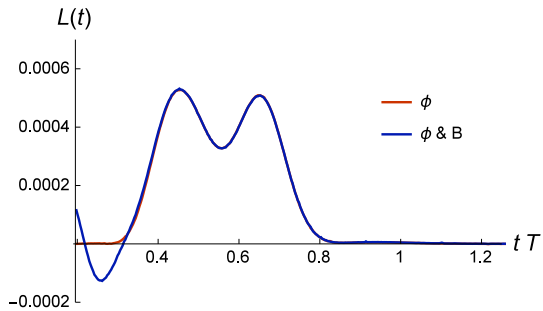}
\end{center}
\caption{\footnotesize 
Geodesic length versus time for the transverse separation $lT=0.31831$. 
Red and blue lines are the graphs for the initial configurations with $B=0$ and both functions present, respectively.
The parameters of the functions are 
$\beta_B=1/6$, ${\cal A}_B=0.1$, $\beta_{\phi}=5/6$ and ${\cal A}_{\phi}=0.5$.
}
\label{L2}
\end{figure}

In the left panel of Fig.\,\ref{L} we have presented the rescaled geodesic length for two points separated by $lT=0.31831$ in the transverse direction. 
Here we consider the evolution of three different initial out-of-equilibrium plasma: 
anisotropic plasma with no non-probe scalar field ($B(0,r)\neq 0$, $\phi(0,r)=0$), isotropic and plasma with non-probe scalar field ($B(0,r)=0$, $\phi(0,r)\neq 0$), and anisotropic plasma with non-probe scalar field (nonzero $B(0,r)$ and $\phi(0,r)$).
From the graphs we can see that for an initial configuration with the same values of $\beta$ and ${\cal A}$ for both functions of the non-probe scalar field and anisotropy, the equilibration time of $L(t)$ mostly comes from the anisotropy function.
As can be seen in Fig.\,\ref{L2}, the non-probe scalar field has noticeable effects on the evolution of $L(t)$ only if it is chosen to have significantly larger values of $\beta$ and/or ${\cal A}$ as compared to the anisotropy function.
These results indicate that the anisotropy function can drive the geometry out of equilibrium much easier than the non-probe scalar field, as viewed by the two-point function.

It is interesting to consider the thermalization time for the first two backgrounds.
The entropy density versus time for these backgrounds has been shown in the right panel of Fig.\,\ref{L}.
It is obvious that the time at which the entropy production ends for these systems is very close.
By choosing $\epsilon$ to be $0.001$, the ratio of $t_{\mathrm{th}}$ for the second system with respect to the first one is obtained about $1.094$, while the same ratio for $t_{\mathrm{L}}$ is about $0.55$.
This means that from the viewpoint of the two-point function, as a nonlocal probe, the first system is much farther away from equilibrium than the second one, while taking the entropy density as an observable which is also sensitive to lower energy scales in the field theory (and equivalently sensitive to the bulk geometry in the deep IR) there is no considerable difference between these systems.
The large difference between the equilibration time of the two-point function on top of these systems may refer to the result we found in \cite{us}, which indicates that the scalar field acts as a probe to the AdS-black hole after the event horizon reaches its equilibrium.

\begin{figure}[htp]
\begin{center}
\includegraphics[width=6.5cm]{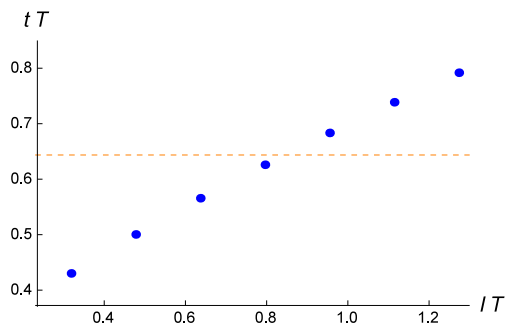}
\end{center}
\caption{\footnotesize 
Equilibration time of the 2-point function versus the separation $l$, for an initial configuration in which both the anisotropy function and scalar field are in the form of the profile (\ref{prof}) with parameters $\beta_B=0.4$, ${\cal A}_B=1$, $\beta_{\phi}=1/3$ and ${\cal A}_{\phi}=1$.
The dashed line indicates the equilibration time of the response of the scalar field.
}
\label{linear}
\end{figure}

Now, we want to compare the time evolution of two different observables:
the normalizable mode of the non-probe scalar field $\phi$, which corresponds to the one-point function of the dual operator in the field theory and therefore is a local observable, and the length of the geodesic connecting two points on the boundary, which gives the correlation between these points in the field theory and hence is a nonlocal probe with a nonzero size.
Choosing the initial anisotropy function and scalar field in the form of the profile (\ref{prof}) with parameters $\beta_B=0.4$, ${\cal A}_B=1$, $\beta_{\phi}=1/3$ and ${\cal A}_{\phi}=1$, we have let the system to evolve and approach its equilibrium.
Then, on top of this time-dependent background, we have calculated the equilibration time of the correlation function between two points on the boundary, and indicated the results for different separations $l$ in Fig.\,\ref{linear} by some points.
From this figure the dependence of the equilibration time on the size of the probe is evident, as is known in the literature (for example \cite{balasubramanian2}), i.e.\;wider probes thermalize later than narrower ones.
We have also shown the equilibration time of the non-probe scalar field response in Fig.\,\ref{linear} by a horizontal dashed line.
It is important to notice that not for all values of $l$ the two-point function equilibrates after the one-point function of the scalar field.

\begin{figure}[htp]
\begin{center}
\includegraphics[width=6.cm]{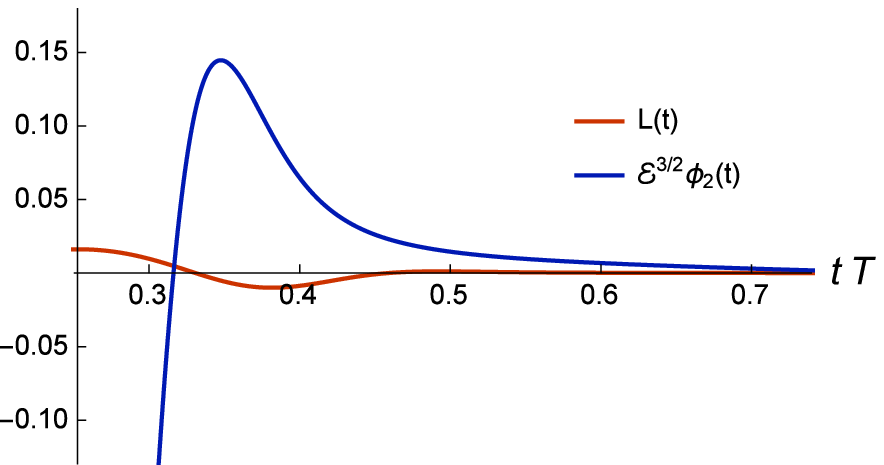}
\hspace{.7cm}
\includegraphics[width=6.cm]{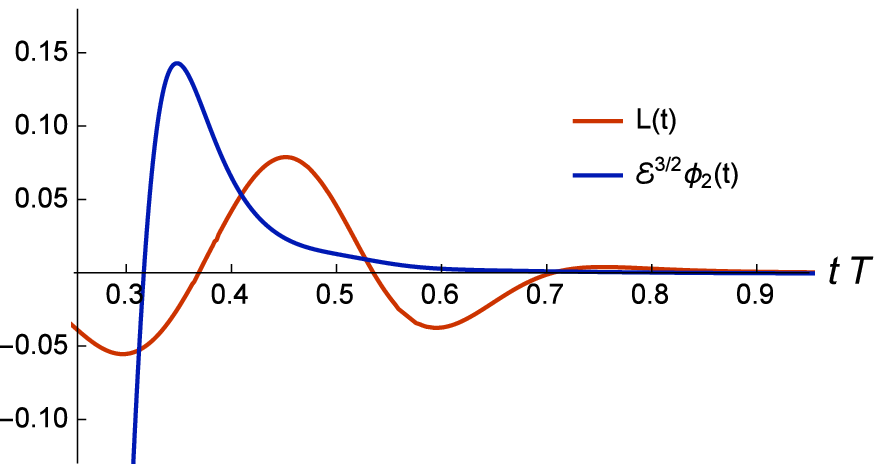}
\end{center}
\caption{\footnotesize 
The parameters of the initial functions in the left panel are $\beta_B=0.4$, ${\cal A}_B=1$, $\beta_{\phi}=1/3$ and ${\cal A}_{\phi}=1$, and in the right panel are $\beta_B=0.7$, ${\cal A}_B=1.2$, $\beta_{\phi}=1/3$ and ${\cal A}_{\phi}=1$.
Red and blue lines are the rescaled geodesic length and the scalar field response, in both graphs.
}
\label{order}
\end{figure}

From Fig.\,\ref{linear} we see that $t_{\mathrm{L}}<t_{\mathrm{eq}}$ for the correlation between two points with the transverse separation $lT=0.31831$.
The time evolution of the two observables for $lT=0.31831$ has been depicted in the left panel of Fig.\,\ref{order}.
In the right panel of this figure the evolution of the two-point function for another initial configuration with parameters $\beta_B=0.7$, ${\cal A}_B=1.2$, $\beta_{\phi}=1/3$ and ${\cal A}_{\phi}=1$ has been shown.
Although in both cases $lT=0.31831$, it is obvious that in the left graph the two-point function equilibrates before the one-point function, while in the right graph the local operator equilibrates first.
One can draw the conclusion that by changing the initial configuration the ordering of the equilibration of local and nonlocal probes can change.
Therefore, the ordering of the equilibration times can be determined only after the exact form of the initial configuration is known.
There is an important issue we should address here.
If there exists only one source for driving the system out of equilibrium, the situation is different.
In this case the ordering of the local and nonlocal equilibration times cannot be reversed by just changing the parameters of the initial configuration.
For example, suppose that for a chosen out-of-equilibrium system which is isotropic from the beginning and the scalar field is in the form of Eq.\,(\ref{prof}) with some parameters $\beta_{\phi}$ and ${\cal A}_{\phi}$, $t_{\mathrm{L}}<t_{\mathrm{eq}}$ for transverse separation $lT=0.31831$.
This relation will always hold, no matter how we change the parameters of the initial profile.
 
\begin{figure}[h]
\begin{center}
\includegraphics[width=6.5cm]{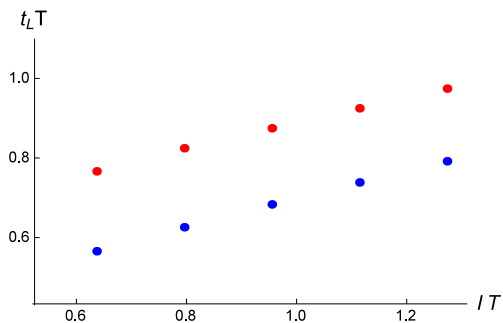}
\end{center}
\caption{\footnotesize 
The equilibration time of the two-point function versus the size $l$ shown with the blue and red dots, respectively, for the following initial configurations. In the first case the anisotropy function and scalar field are in the form of the profile (\ref{prof}) with parameters $\beta_B=0.4$, ${\cal A}_B=1$, $\beta_{\phi}=1/3$ and ${\cal A}_{\phi}=1$, and in the second one $B(0,r)=2\ r^{-4}$ and $\phi(0,r)=0$. 
}
\label{par}
\end{figure}

Another interesting observation from Fig.\,\ref{linear} is the linear increase of the two-point function equilibration time $t_{\mathrm{L}}$ for large enough values of the size $l$.
This feature has also been observed in other systems \cite{nonlocal1,nonlocal4}. 
By checking other initial conditions, we found that this behavior is independent of the initial configuration that we start with.
Moreover, we found that for different initial configurations $t_{\mathrm{L}}$ increases linearly in time with approximately the same slope, at least for the range of the separations $lT$ that we work with.
This behavior can be observed in Fig.\,\ref{par} where we have shown the results for two cases.
Blue dots are the results obtained for the same configuration as the one drawn in Fig.\,\ref{linear} in which both the anisotropy function and non-probe scalar field have the form of the profile (\ref{prof}) with parameters $\beta_B=0.4$, ${\cal A}_B=1$, $\beta_{\phi}=1/3$ and ${\cal A}_{\phi}=1$.
Red dots show the equilibration times for a completely different initial configuration in which no scalar field is present and the anisotropy function is in the form of $B(0,r)=2\ r^{-4}$.
This feature seems to emerge due to the invariance of the initial conditions under the translational symmetry.
Notice that all the initial configurations employed here, are independent of the spatial coordinates of the boundary, i.e.\;$x_L$ and $\bold{x}_T$.
It would be interesting to consider this problem in nontrivial cases where such an invariance is absent, for example in the case of shock wave collisions.

\begin{figure}[h]
\begin{center}
\includegraphics[width=7.cm]{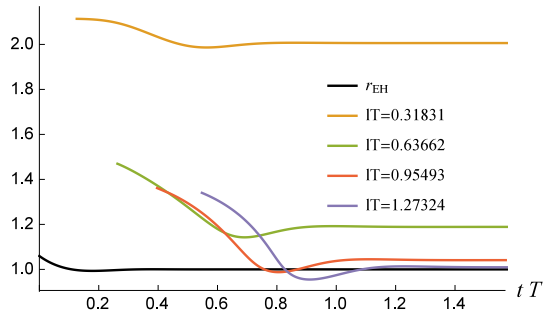}
\end{center}
\caption{\footnotesize 
The $r$ position of the tip of various geodesics corresponding to different transverse separations at the boundary along with the $r$ position of the event horizon (the black curve).
}
\label{varl}
\end{figure}

The evolution of the tip of the geodesics located at $x=0$ is displayed in Fig.\,\ref{varl} for different transverse separations. 
In this figure we also show the position of the event horizon as time proceeds. 
Here the anisotropy function on the initial time slice is chosen to be $B(0,r)=2\ r^{-4}$ while the scalar field has been turned off from the beginning.
As can be seen, for the initial configuration chosen here, the event horizon reaches its equilibrium very early at a time much less than $0.2$. 
However, the two-point functions approach their equilibrium much later.
It is also evident that the time for the equilibration as viewed by these probes, depends on the size of the probe in the field theory and this again supports the idea of faster thermalization at higher energies for strongly-coupled theories.
An interesting phenomenon that can be observed from this figure, is that for sufficiently large boundary separations and after enough amount of time the geodesics cross the event horizon, while at early times they are all outside the horizon.
At sufficiently late times they all approach their equilibrium values which are outside the event horizon. 
It would be interesting to check this result for the case of the extremal surfaces dual to the holographic entanglement entropy in the present background and compare it with the surprising result found in \cite{nonlocal3} for colliding gravitational shock waves. 
They found that for larger separations the geodesics can extend behind the apparent horizon, meaning that the two-point correlation functions probe behind the horizon, while they did not found such a conclusion for the entanglement entropy.
\begin{figure}[h]
\begin{center}
\includegraphics[width=7.cm]{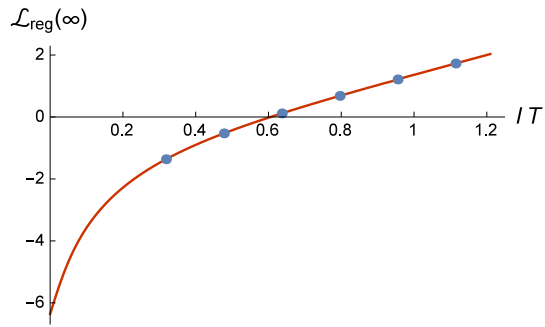}
\end{center}
\caption{\footnotesize 
The regularized geodesic length as a function of $lT$ for the final static case (red line) and the limit of the time-dependent geodesic lengths obtained by the numerical calculation on top of the evolving background (blue dots).
}
\label{static}
\end{figure} 

As the final note we make a comparison between the final equilibrated values of the geodesic lengths calculated so far for the dynamical background with the geodesic lengths obtained directly on top of the static final background.
By passing time, our plasma eventually reaches its equilibrium which is an isotropic plasma with temperature $T$.
In the gravity side this corresponds to an AdS-black hole geometry with the metric
\begin{align}\label{BH} %
 ds^2=2dv dr - r^2\left(1-\frac{(\pi T)^4}{r^4}\right)dv^2+r^2 d\vec{x}^2.
\end{align} %
For this static field theory, one can easily calculate the two-point correlation function between two points separated by an arbitrary length $lT$, numerically, as done in \cite{balasubramanian2}.  
We have sketched the geodesic length for different separations for the static geometry (\ref{BH}) and showed the result by the red line in Fig.\,\ref{static}.
We furthermore show the final equilibrated values of the geodesic lengths on top of the dynamical background, obtained using the numerical calculations explained in Sec. 3.1, by blue dots in this figure. 
As can be seen there is a perfect match between these two calculations which is a sign of the noticeable accuracy of our numerical calculations.
Moreover, this figure shows that the values of $lT$ that we are able to work with numerically for the dynamical case are not actually small, as the blue dots lie in the region that the two-point function decays exponentially with respect to $lT$ which is similar to the behavior at high temperature ($lT\to \infty$), stated in \cite{limits}.
\section*{Acknowledgement}
We thank Michal P. Heller for very useful discussions and also giving lectures at IPM.
H. E. would like to thank Wilke van der Schee for fruitful discussions. M. A. would like to thank School of Physics of Institute for research in fundamental sciences (IPM) for the research facilities and environment. For symbolic GR calculations we used Matthew Headrick's excellent Mathematica package diffgeo.m.

\appendix
\section*{Appendix A} \label{Calculation}
\setcounter{equation}{0}
To solve the Einestein equations (\ref{EOM}), we use pseudo-spectral methods following \cite{Heller2}.
For removing the divergent pieces of the functions $A(v,r)$, $B(v,r)$, $\Sigma(v,r)$ and $\phi(v,r)$ near the boundary, we redefine them as follows
\bse\begin{align}\label{newboundary}
A(v,r)&\rightarrow r^2+ \frac{1}{r} A(v,r),\ \ \ B(v,r)\rightarrow \frac{1}{r^3} B(v,r),\cr
\Sigma(v,r)&\rightarrow r+ \frac{1}{r^2} \Sigma(v,r),\ \ \ \phi(v,r)\rightarrow \frac{1}{r^2} \phi(v,r),\cr
\dot{\Sigma}(v,r)&\rightarrow\frac{r^2}{2}+ \frac{1}{2r^2}\dot{\Sigma}(v,r),\ \ \ \dot{B}(v,r)\rightarrow-2r^3\dot{B}(v,r),\cr
\dot{\phi}(v,r)&\rightarrow \frac{-3}{2r^2}\dot{\phi}(v,r).
\end{align}\ese %
 Then, the following conditions should be satisfied at the boundary by the redefined functions
\bse\begin{align}\label{newboundary1} %
 \Sigma(v,0)&=\Sigma'(v,0)=0,\ \ \ \dot{\Sigma}(v,0)=-\frac{4}{3}{\cal E},\cr
 \dot{B}(v,0)&=B'(v,0),\cr
 \dot{\phi}(v,0)&=\phi'(v,0),\cr
 A(v,0)&=0,\ \ \ A'(v,0)=-\frac{4}{3}{\cal E}.
\end{align}\ese %
Here ${\cal E}$ denotes the energy density which is related to the event horizon of the final static solution as ${\cal E}=\frac{4}{3}r_h^4$.

Notice that we need to have the metric functions on a large enough computational domain for calculation of two-point functions with large separations, since their corresponding geodesics extend deep into the bulk.
We integrated the radial coordinate $z$ ($=\frac{1}{r}$) from the boundary ($z=0$) down to $z=1.6$, far enough behind the final event horizon which has been located at $z=1$.
We found this wide region adequate for the largest $l$ in this paper.
In most of the calculations we chose 60 grid points on the spatial direction of the computational region and set the time steps to be $\delta v=\frac{1}{3000}$.
We found a monotonic growth for the area of the event horizon in all the cases.
This ensures that the abovementioned values provide enough precision for our calculations.

 \end{document}